\documentclass
[10pt, final, conference]{IEEEtran}
\IEEEoverridecommandlockouts
\usepackage[cmex10]{amsmath}
\usepackage{amsfonts}
\usepackage{amssymb}
\usepackage{amscd}
\usepackage{mathrsfs}
\usepackage{afterpage}
\usepackage{graphicx}
\usepackage{cite}

\newcommand{\beq}{\begin{equation}}
\newcommand{\eeq}{\end{equation}}
\newcommand{\beqa}{\begin{eqnarray}}
\newcommand{\eeqa}{\end{eqnarray}}

\begin{document}

\title{Half-Duplex or Full-Duplex Relaying: A Capacity Analysis under Self-Interference }

\author{
\IEEEauthorblockN{Nirmal Shende}
\IEEEauthorblockA{Department of ECE,\\
Polytechnic Institute of NYU\\
Brooklyn, NY 11201\\
Email: nvs235@students.poly.edu}
\and
\IEEEauthorblockN{Ozgur Gurbuz}
\IEEEauthorblockA{Faculty of Engineering and Natural Sciences,\\
Sabanci University\\
Istanbul, TURKEY\\
Email: ogurbuz@sabanciuniv.edu}
\and
\IEEEauthorblockN{Elza Erkip}
\IEEEauthorblockA{ Department of ECE,\\
Polytechnic Institute of NYU\\
Brooklyn, NY 11201\\
Email: elza@poly.edu}
}

\maketitle
\begin{abstract}
In this paper  multi-antenna half-duplex and full-duplex relaying are compared from the perspective of achievable rates. Full-duplexing operation requires additional resources at the relay   such as antennas and  RF chains for self-interference cancellation.  Using a practical model for the residual self-interference,  full-duplex achievable rates and degrees of freedom are computed for the  cases for which the relay  has the same number of antennas or the same number of  RF chains as in the half-duplex case, and compared with their half-duplex counterparts. It is shown that power scaling at the relay  is  necessary to maximize  the degrees of freedom in the full-duplex mode.
\end{abstract}

\section{Introduction}
We consider the problem of communicating from a source  to a destination through a relay, where the destination does not directly receive signals  from the source. Traditional relay systems operate in a half-duplex mode, where the source-to-relay and the relay-to-destination links are kept orthogonal by either frequency division or time division multiplexing. In the full-duplex mode on the other hand, the source and relay  can share a common time-frequency signal-space, so that the relay  can transmit and receive simultaneously over the same frequency band.

Theoretically, half-duplex operation causes loss of spectral efficiency since the capacity achieved in the full-duplex mode can be twice as that of the half-duplex mode. However, full-duplex operation is difficult to implement in practice, because of the significant amount of self-interference observed at the receiving antenna as a result of the signal from the transmitting antenna of the same node. Recently, there has been considerable research interest and promising results in mitigating this self-interference for building practical full-duplex radios \cite{Duarte10,Duarte11,Jain11,Knox12}. The main techniques used to cancel the self-interference are, antenna separation,  a simple passive method, where the self-interference is attenuated due to the path-loss between the transmitting and receiving antennas on the full-duplex node; analog cancellation, an active method, where additional radio frequency (RF) chains are utilized to cancel the interfering signal at the receiving antenna; and digital cancellation, another active cancellation strategy, where the self-interference is removed in the baseband level after the analog-to-digital converter \cite{Duarte11}.

The investigation of the practical benefits of the full-duplex radio and its advantages over half-duplex remains to be an open problem, mainly due to
self-interference experienced. The self-interference due to full-duplex operation can be  theoretically modeled as in \cite{Day12}; however this analysis becomes intractable and requires approximations. Self-interference has also been modeled as an increase in noise floor due to the interfering signal \cite{Aggarwal12}, but this model does not capture the  fact that higher transmit power results in better self-interference mitigation, as it results in better self-interference channel estimates. An upper bound on the the performance of the full-duplex mode can be obtained by assuming perfect self-interference cancellation as in \cite{Barghi12}, but this bound can be overly optimistic in practice. Multi-hop full-duplexing was considered in \cite{Aryafar12},  where
the  signal to noise plus 
interference ratio (SINR) was obtained experimentally for the full-duplex mode. Authors also studied the single cell multiuser scenario for full-duplexing via experiments. In \cite{Riihonen11} Riihonen et. al. provided a comparison of half duplex and full duplex relaying with self-interference, and also discussed the importance of relay power control, but the self-interference model used is simpler than the general model obtained in \cite{Duarte11}.

In this paper, we study a three-node, two hop relay network, where all nodes have multiple antennas and the relay  operates via the decode-and-forward protocol. We compare half-duplex and full-duplex relaying while we explicitly consider the residual self-interference using the  model in \cite{Duarte11} for full-duplex mode. This  is an analytically tractable, experimentally developed model and it incorporates the effect of the relay transmit power on the self-interference. Furthermore, it also encompasses all the other self-interference models in the literature as special cases. We formulate, calculate and compare the achievable rates and degrees of freedom, when the relay  operates in the half-duplex and full-duplex modes. In order to keep our comparisons fair, we consider two different scenarios as in \cite{Aggarwal12}, where different radio resources allocated to the relay  in half-duplex mode and full-duplex mode are kept constant. In the \textit{antenna conserved} scenario, it is assumed that the total number of antennas used by the relay  is kept the same for the half-duplex and full-duplex modes, and in the \textit{RF chain conserved} scenario, the total number of up-converting and down-converting RF chains is kept the same for the half-duplex and full-duplex modes. In either scenario, for the full-duplex mode, we show the benefits of relay power control for maximizing the achievable rates and the degrees of freedom. Whether the full-duplex mode or the half-duplex mode performs better depends on the scenario  investigated (antenna conserved or RF chain conserved), the number of antennas at the nodes, the operating signal to noise ratio (SNR), and level of interference suppression achieved.

The rest of the paper is organized as follows. In Section \ref{sec:CM}, the system model is introduced. In Sections \ref{sec:HD} and \ref{sec:FD}, the half-duplex and full-duplex relaying modes are studied via achievable rates under finite SNR and via degrees of freedom (DoF) under high SNR. In Section \ref{sec:comp}, numerical results are presented for both performance metrics, comparing half-duplex and full-duplex relaying in the antenna and RF chain conserved scenarios. Section \ref{sec:Disc} provides our conclusions.

\begin{figure}[!h]
\centering
\includegraphics[scale=0.40]{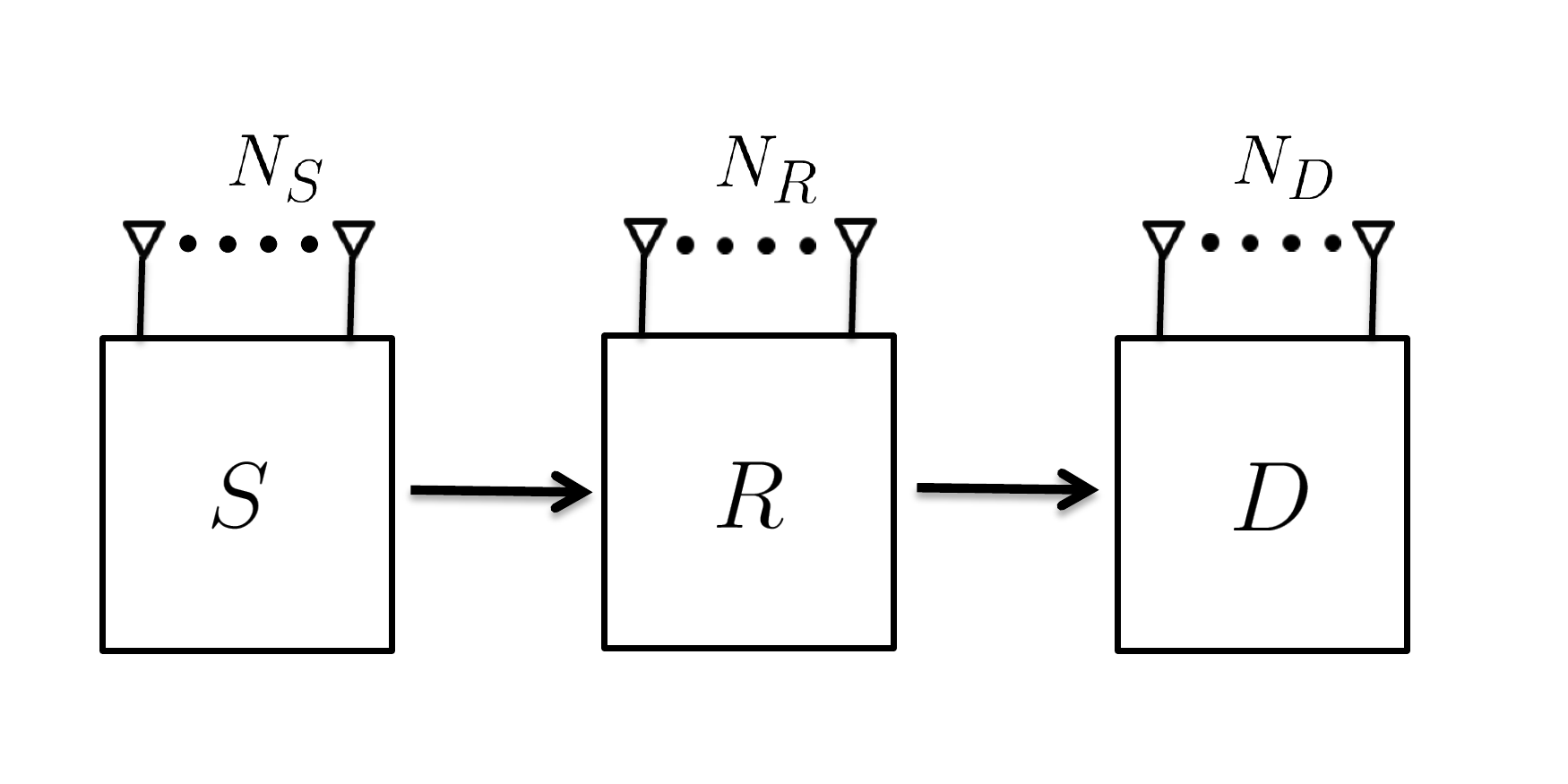}
\caption{A three node, two hop relay network.}
\label{fig:relay1}
\end{figure}

\section{System Model}
\label{sec:CM}
We consider a three node network with a source, a relay  and a destination  as shown in Figure \ref{fig:relay1}. The source and the destination  have $N_S$ and $N_D$ antennas respectively, and cannot communicate directly. The relay, while operating in the half-duplex mode, has a total $N_R$ antennas, corresponding to $N_R$ up-converting radio chains for the transmission and $N_R$ down-converting radio chains for the reception, resulting in total number of radio chains as $2N_R$. The number of antennas and RF chains in the full duplex scenario will depend on whether we have the antenna conserved or RF chain conserved scenario, and will be discussed in Section \ref{sec:FD}. 

Let $P_S$ denote the average power at the source, $P_R$ denote the average power constraint at the relay, $\sigma_R^2$ and $\sigma_D^2$ denote the average power of additive white Gaussian noise (AWGN) at the relay and destination respectively. The path-loss over the source-to-relay link and the relay-to-destination link are given by $K d_1^{\gamma}$ and $K d_2^{\gamma}$ respectively, where $d_1$ and $d_2$ are the distances between the corresponding nodes, $K$ is a constant summarizing the effects of the propagation medium, communication frequency and antenna parameters, and $\gamma$ is the path loss exponent, which depends upon the communication environment \cite{Rappaport}. The source-relay, and relay-destination channels are assumed to have independent Rayleigh fading with the channel state information available at the receiver (CSIR), but not at the transmitter. The communication is assumed to last long enough so that ergodic rates can be achieved.

We let matrix $\mathcal{H}$ denote the random fading channel coefficients between the source and the relay, and  matrix $\mathcal{G}$ denote the random fading channel coefficients between  the relay and the destination. Because of the Rayleigh fading assumption, the entries of $\mathcal{H}$ and $\mathcal{G}$ are assumed to follow the circularly symmetric complex Gaussian distribution with unit variance. The dimensions of $\mathcal{H}$ and $\mathcal{G}$ depend upon the mode of operation and scenario, i.e., half-duplex or full-duplex, and also on the antenna conserved or the RF conserved scenario, and will be discussed in detail in Sections \ref{sec:HD} and \ref{sec:FD}.


\section{Half Duplex Relaying}
\label{sec:HD}

\subsection{Channel Model}
When the relay  operates in the half-duplex mode, it can either transmit or receive at any given time, where in general, a fraction $\tau$ of time is allocated for reception and the remaining fraction of $(1-\tau)$ is used for transmission. While receiving from the source , the relay utilizes its $N_R$ antennas for the receive diversity, and when transmitting to the destination, it transmits using the same $N_R$ antennas, employing transmit diversity.

Let $\mathbf{x}_S$ and $\mathbf{x}_R$ be the symbols transmitted by the source and relay respectively, and $\mathbf{y}_R$ and $\mathbf{y}_D$ denote the signals received by the relay and the destination  respectively. Also, $\mathbf{w}_R$ and $\mathbf{w}_D$ are the AWGN terms at the relay and and destination.

The channel model for the half-duplex mode is given as,
\begin{eqnarray}
\mathbf{y}_R=\frac{1}{\sqrt{K d_1^{\gamma}}}\mathcal{H}\mathbf{x}_S+\mathbf{w}_R, \\
\mathbf{y}_D=\frac{1}{\sqrt{K d_2^\gamma}}\mathcal{G}\mathbf{x}_R+\mathbf{w}_D.
\end{eqnarray}
Here, $\mathcal{H}\in\mathbb{C}^{N_R \times N_S}$, $\mathcal{G} \in \mathbb{C}^{N_D \times N_R}$, $\mathbf{x}_S \in \mathbb{C}^{N_S \times 1}$, $\mathbf{y}_R, \mathbf{w}_R \in \mathbb{C}^{N_R \times 1}$, $\mathbf{x}_R \in \mathbb{C}^{N_R \times 1}$, $\mathbf{y}_D, \mathbf{w}_D \in \mathbb{C}^{N_D\times 1}$.

The average SNR at the relay and destination can be obtained as
\beq
\text{SNR}_R=\frac{P_S}{K d_1^{\gamma}\sigma^2_R},
\label{eq:SNR_R}
\eeq
\beq
\text{SNR}_D=\frac{P_R}{K d_2^\gamma\sigma^2_D}.
\label{eq:SNR_D}
\eeq

\subsection{Achievable Rate}
In this subsection, we derive the achievable rate considering finite SNR levels.  Given that the source-relay link is active for $\tau$ of time and the relay-destination link is used $(1-\tau)$ of time, to ensure the power constraints $P_S$ and $P_R$ are satisfied,  $\mathbf{x}_S$ and $\mathbf{x}_R$ can be scaled such that $\mathbb{E}[\mathbf{x}_S^*\mathbf{x}_S]=\frac{P_S}{ \tau}$ and $\mathbb{E}[\mathbf{x}_R^*\mathbf{x}_R]=\frac{P_R}{(1-\tau)}$, where $\mathbf{x}_S^*$ denotes the Hermitian of the vector $\mathbf{x}_S$.

The average rate achievable on the source-relay link is then obtained by \cite{Tse05},
\beq
R_{SR}^{HD}=\tau\mathbb{E}_{\mathcal{H}}\left[\log \det \left(\mathcal{I}+\frac{\text{SNR}_R}{\tau N_S}\mathcal{H}\mathcal{H}^*\right)\right],
\eeq
where $\mathcal{I}$ is the identity matrix. Similarly, the rate achievable over the relay-destination link is given by
\beq
R_{RD}^{HD}=(1-\tau)\mathbb{E}_{\mathcal{G}}\left[\log \det \left(\mathcal{I}+\frac{\text{SNR}_D}{(1-\tau)N_R}\mathcal{G}\mathcal{G}^*\right)\right]. \nonumber
\eeq
By optimizing over $\tau$, the end-to-end average achievable rate for half-duplex relaying can be found as
\beq
R_{HD}=\max_{0\le\tau \le 1}\min\left\{R_{SR}^{HD},R_{RD}^{HD}\right\}.
\label{eq:C_HD}
\eeq
\subsection{Degrees of Freedom}
Degrees of Freedom (DoF) analysis characterizes the achievable rate at high SNR. For a point to point multi-input multi-output (MIMO) AWGN Rayleigh fading channel with $N_S$ antennas at the source  and $N_D$ antennas at the destination, the largest DoF is given by\cite{Tse05}
\beq
\text{DoF}=\lim_{\text{SNR} \to \infty}=\frac{R(\text{SNR})}{\log(\text{SNR})}=\min\left\{N_D,N_S\right\},
\eeq
where $R({\text{SNR})}$ is the largest rate achieved at SNR.

For the DoF of the relay network, since there are two SNR terms involved in the expression of $R_{HD}$, we assume that the relay scales its power relative to the source's power according to
\beq
P_R=P_S^c, \,\,\,  \, 0<c\le1.
\eeq
Then,  the DoF of the relay network in the half-duplex mode is given by 
\begin{align}
\text{DoF}_{HD}=\sup_{0<c\le1}\lim_{\substack{\text{SNR}_R \to \infty  \\ \text{SNR}_D =\text{SNR}_R^c}}&\frac{R_{HD}}{\log(\text{SNR}_R)} \\
=\max_{\substack{ 0<\tau<1 \\ 0<c\le 1}}\min \{&\tau\min\{N_S,N_R\},\nonumber\\
&c(1-\tau)\min\{N_R,N_D\}\} \label{eq:C}\\
= \max_{0<\tau<1} \min &\{\tau N_S,\tau N_R,\nonumber\\
&(1-\tau)N_R,(1-\tau
)N_D\}.
\end{align}
Note that in (\ref{eq:C}), setting $c=1$ maximizes the $\text{DoF}_{HD}$.

Depending on the values of $N_S$, $N_R$, and $N_D$, and using optimal $\tau$ denoted as $\tau_{opt}$, we obtain following DoF values for the half-duplex mode as shown in the Table \ref{tab:DOFHD}.

\begin{table}[!h]
\centering
\begin{tabular}{| c | c | c |}

\hline
   & $\tau_{opt}$ & $\text{DoF}_{HD}$ \\ [8pt] \hline
  $N_R \le \min\{{N_s,N_D}\}$ & $\frac{1}{2}$ & $\frac{N_R}{2}$ \\ [8pt] \hline
  $N_R \ge \max\{{N_s,N_D}\}$ & $\frac{N_D}{N_D+N_S}$ & $\frac{N_SN_D}{N_D+N_S}$ \\ [8pt] \hline
  $N_S \le N_R \le N_D$ & $\frac{N_R}{N_R+N_S}$ & $\frac{N_RN_S}{N_R+N_S}$ \\ [8pt]\hline
  $N_D \le N_R \le N_S$ & $\frac{N_D}{N_D+N_R}$ & $\frac{N_RN_D}{N_R+N_D}$ \\ [8pt] \hline

\end{tabular}
\caption{Degrees of Freedom for the Half-duplex Mode}
\label{tab:DOFHD}
\end{table}

\section{Full Duplex Relaying}
\label{sec:FD}
In the full-duplex mode, the relay  is able to receive and transmit simultaneously, however, it is subjected to the self-interference. To maintain the causality, the relay transmits $(i-1)^{\text{th}}$ packet while it receives the $i^{\text{th}}$ packet. The relay can use several cancellation techniques to mitigate the self-interference \cite{Duarte11}. The simplest self-interference cancellation is the passive one, obtained by the path-loss due to the separation between the transmitting and the receiving antennas. Additional sophisticated techniques, active analog cancellation and active digital cancellation reduce the self-interference further. In analog cancellation, the relay  uses the estimate of the channel between the transmitting and receiving antennas to subtract the interfering signal at the RF stage. Hence, additional RF chains are required for implementing the analog cancellation. In the digital cancellation, the self-interference is canceled in the baseband.

\subsection{Channel Model}
We assume that the relay  allocates $r$ of its antennas for reception and $t$ antennas for transmission in the full-duplex mode. In the {antenna conserved} scenario, the total number of antennas used by the relay  are the same as those used in the half duplex mode. Then, for full-duplex mode with $r$ antennas used for reception, $t=(N_R-r$) antennas can be used for  transmission. For the {RF chain conserved} scenario, if  $r$ antennas are used for  reception, then in addition to the $r$ down-converting RF chains, $r$ up-converting RF chains are necessary for  active analog cancellation. Remaining $(2N_R-2r)$ RF chains can be used for the up-conversation for the transmission, resulting in the number of the transmit antennas as $(2N_R-2r)$. Note that in the RF chain conserved scenario, the relay uses a total of $(2N_R-r)$ antennas, which is larger than $N_R$, the number of antennas in the half-duplex case. This is justified as the cost of an RF chain is typically the dominant factor \cite{Aggarwal12}.

The channel model in this case is expressed by
\begin{align}
\mathbf{y}_R&=\frac{1}{\sqrt{K d_1^{\gamma}}}\mathcal{H}\mathbf{x}_S+\mathbf{w}_R+
\mathbf{i}(\tilde{P}_R), 
\label{eq:Y_R}\\
\mathbf{y}_D&=\frac{1}{\sqrt{K d_2^\gamma}}\mathcal{G}\mathbf{x}_R+\mathbf{w}_D,
\end{align}
where, $\mathcal{H}\in\mathbb{C}^{r \times N_S}$, $\mathcal{G} \in \mathbb{C}^{N_D \times t}$, $\mathbf{x}_S \in \mathbb{C}^{N_S \times 1}$, $\mathbf{y}_R, \mathbf{w}_R, \mathbf{i}(\tilde{P}_R) \in \mathbb{C}^{r \times 1}$, $\mathbf{x}_R \in \mathbb{C}^{t \times 1}$, $\mathbf{y}_D, \mathbf{w}_D \in \mathbb{C}^{N_D\times 1}$.  Here $\mathbf{x}_S$ and $\mathbf{x}_R$ denote the symbols transmitted by the source and relay, and $\mathbf{y}_R$ and $\mathbf{y}_D$ denote the signals received by the relay and the destination, respectively. Also, $\mathbf{w}_R$ and $\mathbf{w}_D$ are the AWGN terms at the relay and and destination. The additional term $\mathbf{i}(\tilde{P}_R)$ in (\ref{eq:Y_R}) represents the residual self-interference at the relay, with average power $I(\tilde{P}_R)$ per receiving antenna, where $\tilde{P}_R \le P_R$ is the average transmit power of the relay. 

From the experimental characterization of the self-interference in \cite{Duarte11}, the average power of the residual self-interference is modeled as

\beq
I(\tilde{P}_R)=\frac{\tilde{P}_R^{(1-\lambda)}}{\beta \mu^\lambda},
\label{eq:SIC}
\eeq
the parameter $\beta$ is the interference suppression factor due to passive cancellation, and parameters $\mu$ and $\lambda$ depend on the self-interference cancellation technique  \cite{Duarte11}.

Then, the average SINR at the relay  is given by
\beq
\text{SINR}_R=\frac{P_S}{K d_1^{\gamma}(I(\tilde{P}_R)+\sigma^2_R)}=\frac{P_S}{K d_1^{\gamma}\left(\frac{\tilde{P}_R^{(1-\lambda)}}{\beta \mu^\lambda}+\sigma^2_R\right)},
\label{eq:SINRR}
\eeq

and the average SNR at the destination is given as

\beq
\text{SNR}_D=\frac{\tilde{P}_R}{K d_2^\gamma\sigma^2_D}.
\eeq

\subsection{Achievable Rate}
We first derive the achievable rate considering finite SNR levels. Let $ R_{SR}^{FD}$ denote the average rate from the source  to the relay. Using (\ref{eq:Y_R}) $ R_{SR}^{FD}$ is given as,

\begin{equation}
R_{SR}^{FD}=\mathbb{E}_{\mathcal{H}}\left[\log \det \left(\mathcal{I}+\frac{\text{SINR}_R}{N_S}\mathcal{H}\mathcal{H}^*\right)\right].
\label{eq:R_FD_SR}
\end{equation}

 Similarly, $ R_{RD}^{FD}$, the average rate from the relay  to the destination  is obtained as,

\begin{equation}
R_{RD}^{FD}=\mathbb{E}_{\mathcal{G}}\left[\log \det \left(\mathcal{I}+\frac{\text{SNR}_D}{t}\mathcal{G}\mathcal{G}^*\right)\right].
\label{eq:R_FD_RD}
\end{equation}
We recall that, $t=(N_R-r)$ for the antenna conserved case and $t=(2N_R-2r)$ for the RF chain conserved case. We assume that the relay  can optimally allocate the number of receive antennas to maximize the average rate achievable from the source  to the destination.
Furthermore, depending upon the average SINR at the relay  and SNR at the destination, the excess power at the relay  can have a negative impact on the achievable rate due to increased self-interference. Note that
from (\ref{eq:SINRR}), the SINR at the relay  decreases as the relay power $\tilde{P}_R$ increases if the  source power $P_S$ is held constant. Thus, with the increase in $\tilde{P}_R$ for a constant $P_S$, the source-to-relay channel rate decreases while the relay-to-destination channel rate increases.

Therefore the achievable rate for the full duplex relaying can be written as 
\beq
R_{FD}=\max_{ \substack{ 0<r<N_R \\   \tilde{P}_R \le P_R}}\min\left\{R_{SR}^{FD},R_{RD}^{FD}\right\}.
\label{eq:C_FD}
\eeq

\begin{figure}[hbt]
\includegraphics[scale=0.45]{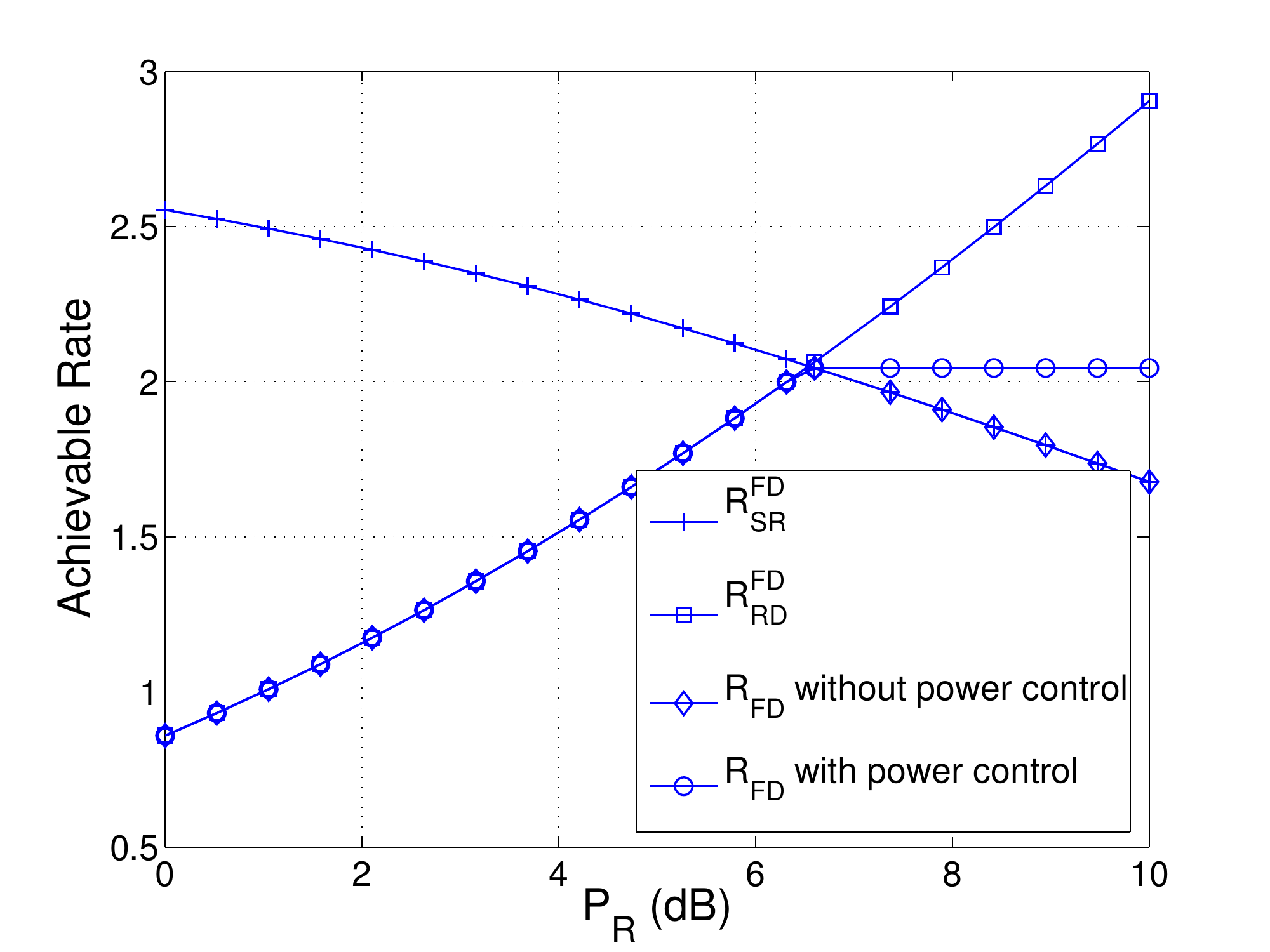}
\caption{Achievable rates for the full-duplex relaying as a function of $P_R$, $P_S$= 10 dB $\lambda=0.2$, $\beta=38$ dB, $\mu=13$ dB, $\sigma^2_R=\sigma^2_D=10^{-5}$, $K d_1^{\gamma}=K d_2^\gamma=50$ dB.}
\label{fig:FD_PC}
\end{figure}

The importance of  relay power control,  also proposed in \cite{Riihonen11}, is illustrated in Figure \ref{fig:FD_PC}, for the antenna conserved full duplex relaying with  $N_S=N_D=1$, $N_R=2$. Here  $R_{SR}^{FD}$ is given in (\ref{eq:R_FD_SR}), $R_{RD}^{FD}$  is given in (\ref{eq:R_FD_RD}), $R_{FD}$ is given in (\ref{eq:C_FD}) with and without power control ($\tilde{P}_R=P_R$).
The optimal relay power level $\tilde{P}_R$  for the above antenna configuration can be computed by solving

\beq
\frac{P_S}{K d_1^{\gamma}\left(\frac{\tilde{P}_R^{(1-\lambda)}}{\beta \mu^\lambda}+\sigma^2_R\right)}=\frac{\tilde{P}_R}{K d_2^\gamma \sigma_D^2}.
\eeq

\subsection{Degrees of Freedom}
Similar to the half-duplex case, the DoF of a full-duplex relay network can be found as 
\beq
\text{DoF}_{FD}=\sup_{0<c\le1}\lim_{\substack{\text{SNR}_R \to \infty \\ \text{SNR}_D =\text{SNR}_R^c}}\frac{R_{FD}}{\log(\text{SNR}_R)},
\eeq
where $\text{SNR}_R$  and $\text{SNR}_D$ are as defined in (\ref{eq:SNR_R}) and (\ref{eq:SNR_D}) respectively.
In order to control the self-interference, the relay  scales its power as compared to the source's power through 
\beq
P_R=P_S^c, \,\,\,  \, 0<c\le1.
\eeq

Then the achievable DoF in the full-duplex mode can be computed as

\begin{align}
\text{DoF}_{FD}=\max_{\substack{0<r<N_R \\ 0<c\le 1}}\min \{&(1-c(1-\lambda))\min\{N_S,r\},\nonumber\\
&c\min\{t,N_D\}\} \nonumber\\
=\max_{\substack{0<r<N_R \\ 0<c\le 1}}\min \{&(1-c(1-\lambda))N_S,(1-c(1-\lambda))r,\nonumber\\
&ct,cN_D\}.
\label{eq:DoF_FD}
\end{align}
Here, $t=(N_R-r)$ for the antenna conserved and $t=(2N_R-2r)$ for the RF chain conserved case. 

In Section \ref{subsec:DoF}, we will explicitly compute the $\text{DoF}_{FD}$ for $N_S=N_D$ case and compare it with $\text{DoF}_{HD}$.

\section{Comparison of Half-Duplex and Full-Duplex Relaying}
\label{sec:comp}

\subsection{Achievable Rate}
In order to compare the achievable rates for finite SNR values, we examine an example case with $N_S=1, N_R=2$ and $N_D=1$.

In the antenna conserved scenario, the only choice for the number of relay receive antennas is $r=1$, resulting in $t=N_R-r=1$.  In the RF chain conserved case, similarly we set $r=1$, and we have $t=2N_R-2r=2$.

\begin{figure}
\includegraphics[scale=0.45]{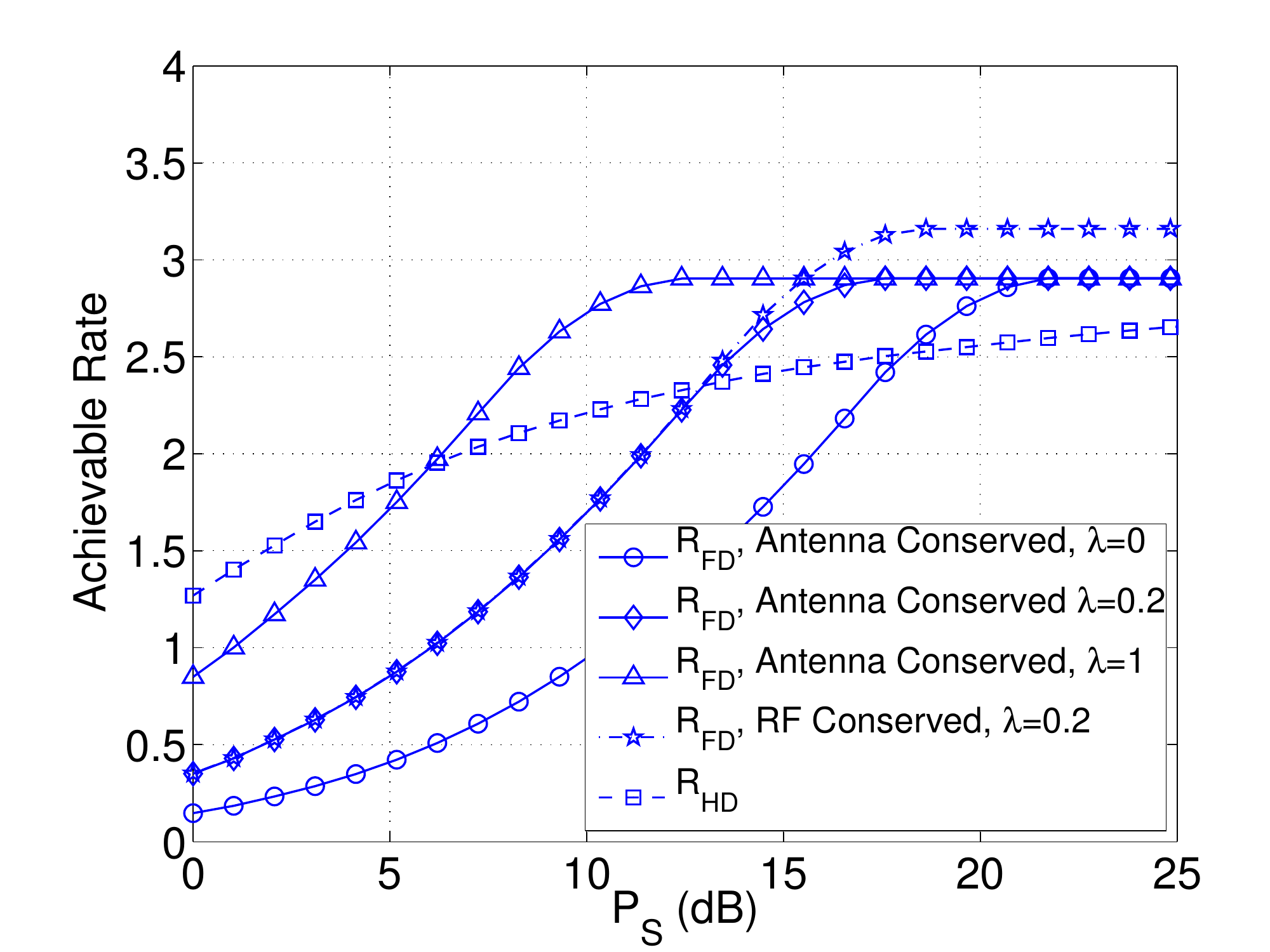}
\caption{Achievable rates for the half-duplex and full-duplex relaying as a function of  $P_S$, $P_R=10$ dB,  $\beta=38$ dB, $\mu=13$ dB, $\sigma^2_R=\sigma^2_D=10^{-5}$, $K d_1^{\gamma}=K d_2^\gamma=50$ dB.}
\label{fig:Ps_lamb}
\end{figure}

Figure \ref{fig:Ps_lamb} depicts the achievable rates for various values of $\lambda$ as a function of $P_S$, when $P_R$ is held constant. We recall from (\ref{eq:SIC}) that $\lambda$ determines the amount of residual self-interference. As expected, the performance of the full-duplex mode improves as the value of $\lambda$ increases, since with higher $\lambda$ better interference suppression is achieved. Also, the RF chain conserved case performs similar to  the antenna conserved case at low $P_S$,  while its performance surpasses the antenna conserved case at higher $P_S$. This is due to the higher number of transmit antennas relay  can support in the RF chain conserved scenario.

\begin{figure}
\includegraphics[scale=0.45]{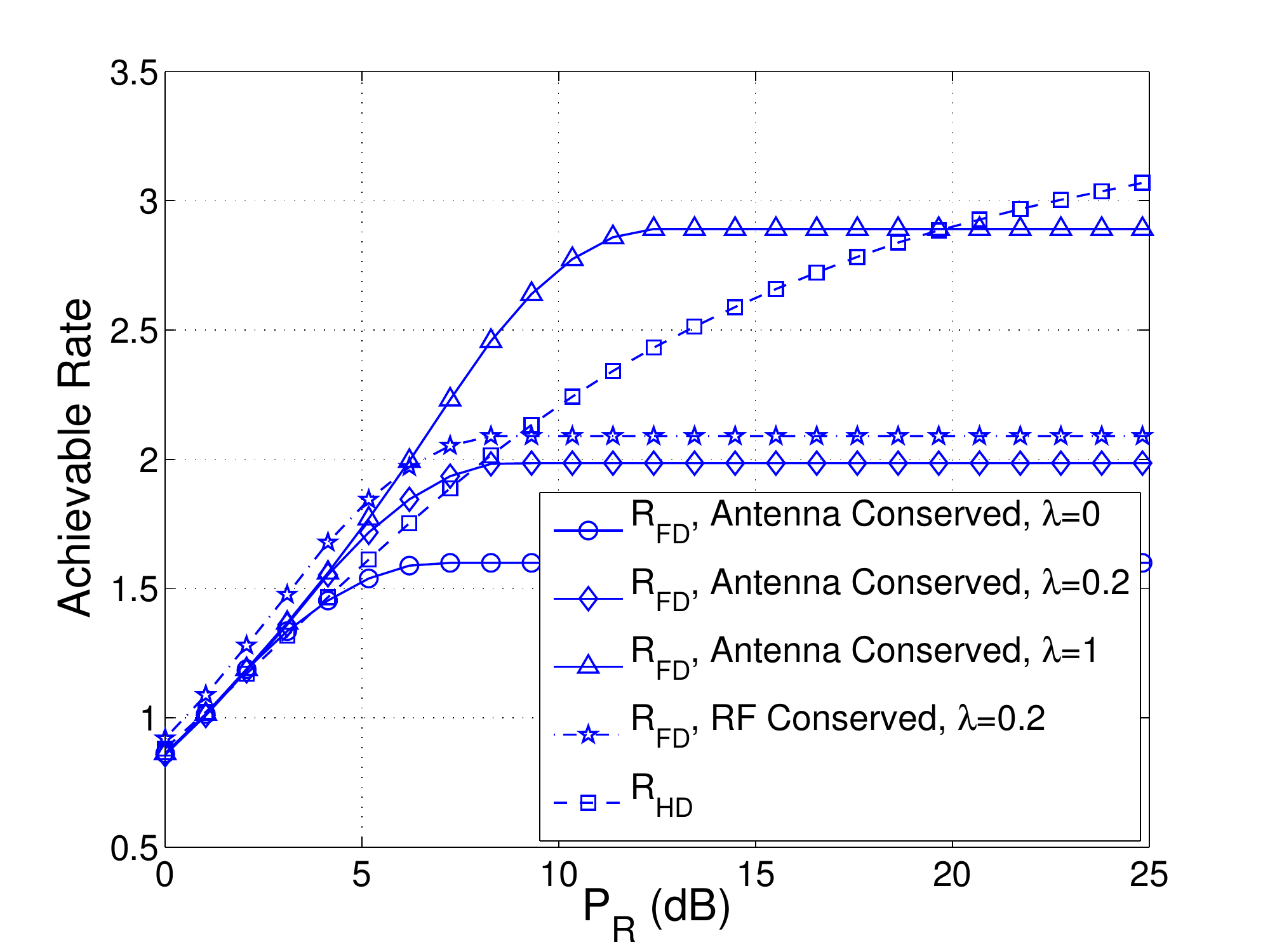}
\caption{Achievable rates for the half-duplex and full-duplex relaying as a function of $P_R$, $P_S=10$ dB,  $\beta=38$ dB, $\mu=13$ dB, $\sigma^2_R=\sigma^2_D=10^{-5}$, $K d_1^{\gamma}=K d_2^\gamma=50$ dB.}
\label{fig:P_R_lamb}
\end{figure}

Figure \ref{fig:P_R_lamb} compares the achievable rate for various values of $\lambda$ as a function of $P_R$, when $P_S$ is held constant. Along with similar observations as in Figure \ref{fig:Ps_lamb}, it can be seen that as the relay's power is increased, the half-duplex rate increases, but the full-duplex rate reaches a plateau  due to the power control to limit the self-interference. Hence for fixed $P_S$, at the higher relay power levels, the half-duplex mode performs better than the full duplex mode, since it can utilize the excess power at the relay  through optimizing $\tau$ to maximize the achievable rate.


\subsection{Degrees of Freedom}
\label{subsec:DoF}
To compare the half-duplex and full-duplex DoF, we consider a special case when the source and destination have same number of antennas, i.e.,
 $N_S$=$N_D$=$N$, we assume  $N$ and $N_R$ are even, then from the Table \ref{tab:DOFHD}
\begin{equation}
\text{DoF}_{HD}=\min\left\{\frac{N}{2},\frac{N_R}{2}\right\}.
\end{equation}

For the full-duplex antenna conserved case with $N_S$=$N_D$=$N$,  (\ref{eq:DoF_FD}) can be written as
\begin{IEEEeqnarray}{rCll}
\text{DoF}_{FD}^{AC}&=&\max_{\substack{1\le r < N_R \\  0<c\le 1}}\min \{&(1-c(1-\lambda))N,(1-c(1-\lambda))r,\nonumber\\
&&\hbox{}&c(N_R-r),cN) \nonumber \} \\
&\le &\max_{0<c\le 1}\min \{(&1-c(1-\lambda))N,cN\} \label{eq:DoF1}\\
&=&\frac{N}{2-\lambda}. \label{eq:DoF2}
\end{IEEEeqnarray}

Since the minimum of two terms in (\ref{eq:DoF1}) is maximized when both are equal, (\ref{eq:DoF2}) can be obtained by setting $c=\frac{1}{2-\lambda}$.

Similarly,
\begin{eqnarray}
\text{DoF}_{FD}^{AC} &\le& \max_{\substack{1\le r < N_R \\ 0<c\le 1}}\min \{(1-c(1-\lambda))r,c(N_R-r)\},\nonumber\\
&=&\frac{N_R}{2(2-\lambda)} \label{eq:DoF3},
\end{eqnarray}
where (\ref{eq:DoF3}) is obtained by setting $c=\frac{1}{2-\lambda}$ and $r=\frac{N_R}{2}$. 

From (\ref{eq:DoF2}) and  (\ref{eq:DoF3}), we can write
\begin{equation}
\text{DoF}_{FD}^{AC} \le \frac{1}{2-\lambda}\min\left\{N,\frac{N_R}{2}\right\}.
\label{eq:DoF4}
\end{equation}

Equality in (\ref{eq:DoF4}) can be achieved when $c=\frac{1}{2-\lambda}$ and $r=\frac{N_R}{2}$, leading to

\begin{equation}
\text{DoF}_{FD}^{AC} =\frac{1}{2-\lambda}\min\left\{N,\frac{N_R}{2}\right\}.
\end{equation}

Similarly, for the RF chain conserved scenario we have, 
\begin{equation}
\text{DoF}_{FD}^{RC}=\frac{1}{2-\lambda}\min\left\{N,\left\lfloor\frac{2N_R}{3}\right\rfloor\right\}.
\label{eq:DoF_FD_RC}
\end{equation}

\begin{figure}
\includegraphics[scale=0.45]{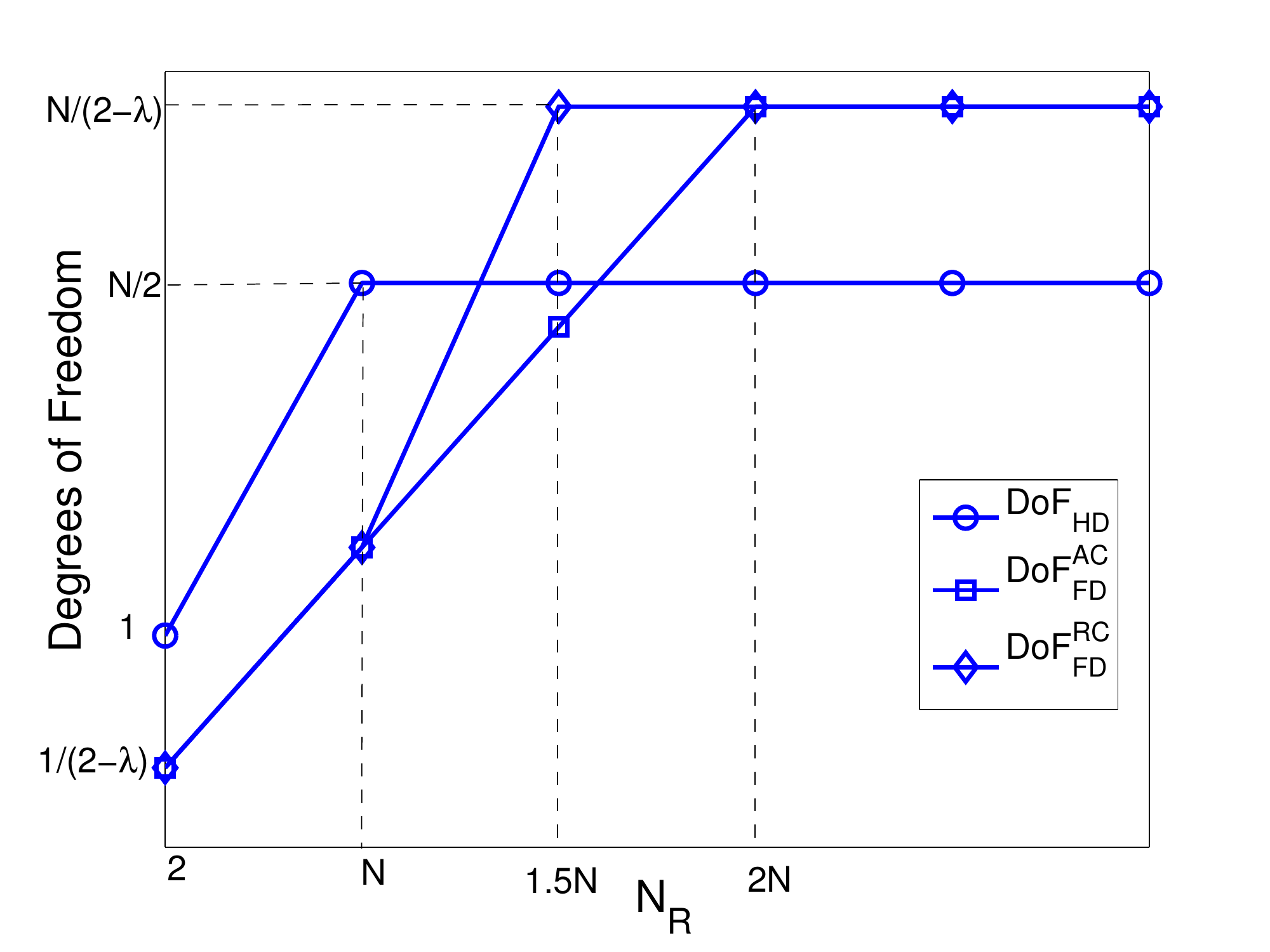}
\caption{The degrees of freedom for the half-duplex and full-duplex relay channel under antenna conserved and RF conserved scenario, $N_S$=$N_D$=$N$. This figure is drawn for $N=4$.}
\label{fig:DoF}
\end{figure}

The degrees of freedom in the half-duplex mode and full-duplex mode for the antenna conserved  and RF chain conserved scenarios are plotted in Figure \ref{fig:DoF}. As seen from the figure, when $N_R$ is small, the half-duplex mode performs better than the full-duplex mode, and the situation is reversed when $N_R$ gets larger, with the RF conserved scenario always dominating the antenna conserved one. The actual cross-over point depends on the number of respective antennas and the self-interference suppression factor $\lambda$.

\section{Conclusions and Discussion}
\label{sec:Disc}
In this paper, we have compared half-duplex and full-duplex relaying, by taking self-interference into account with a general and realistic  model. In our comparisons, we have taken the  number of half-duplex antennas and RF chains as a benchmark and considered two full-duplex scenarios: the antenna conserved and the RF chain conserved.  In order to limit the self-interference in the full-duplex mode, the relay  should exhibit power control, where it does not utilize its full power beyond a certain threshold. Hence, if the relay-destination link has much higher SNR than the source-relay link, then it is recommended to use the half-duplex mode, since it can utilize the higher power at the relay through asymmetrical fractional use of these two links. When the SNRs of these two links are comparable or the source-relay link is weaker, it is advisable to use the full-duplex mode. At high SNR, the full-duplex mode gives higher degrees of freedom when both source and  relay powers are increased with appropriate scaling and relay  has sufficiently large number of antennas as compared to the source and destination. If the relay has smaller number of antennas, the half-duplex mode has higher degrees of freedom, promising higher capacity.

While  we have considered employing
different antennas for the reception and transmission
operations in the full-duplex mode to ensure passive interference cancellation as in \cite{Duarte11}, Knox  \cite{Knox12} presents a  full-duplex design using a single circularly polarized antenna. Such a design would make the full-duplex
mode even more attractive, since we do not need to allocate
different antennas for receiving and transmitting, and will be the subject of a future study. 

\bibliographystyle{IEEEtran}
\bibliography{IEEEabrv,CiSS_2013}

\end{document}